\documentclass[prb,twocolumn,showpacs,preprintnumbers,amsmath,amssymb]{revtex4-1}

\usepackage{graphicx}
\usepackage{dcolumn}
\usepackage{bm}

\begin{document}

\preprint{APS-Draft}

\title{An Experimental and  Semi-Empirical Method to Determine the Pauli-Limiting Field in Quasi 2D Superconductors as applied to $\kappa$-(BEDT-TTF)$_2$Cu(NCS)$_2$: Strong Evidence of a FFLO State}

\author{C. C. Agosta}
\email{cagosta@clarku.edu}
\homepage{http://physics.clarku.edu/~cagosta}
\author{Jing Jin}
\author{W. A. Coniglio}
\author{B. E. Smith}
\author{K. Cho}
\author{I. Stroe}
\author{C. Martin}
\affiliation{Physics Dept., Clark University, 950 Main St., Worcester MA 01610, USA}
\author{S. W. Tozer}
\author{T. P. Murphy}
\author{E. C. Palm}
\affiliation{The National High Magnetic Field Laboratory, 1800 E. Paul Dirac Drive, Tallahassee, FL 32310, USA}
\author{J. A. Schlueter}
\affiliation{Materials Science Division, Argonne National Laboratory, 9700 South Cass Ave. Argonne, IL 60439 USA}
\author{M. Kurmoo}
\affiliation{The Royal Institution, 21 Albemarle Street, London W1X 4BS, United Kingdom}

\date{\today}

\begin{abstract}
We present upper critical field data for  $\kappa$-(BEDT-TTF)$_2$Cu(NCS)$_2$ with the magnetic field close to parallel and parallel to the conducting layers.  We show that we can eliminate the effect of vortex dynamics in these layered materials if the layers are oriented within 0.3 degrees of parallel to the applied magnetic field.  Eliminating vortex effects leaves one remaining feature in the data that corresponds to the Pauli paramagnetic limit ($H_p$). We propose a semi-empirical method to calculate the $H_p$ in quasi 2D superconductors.  This method takes into account the energy gap of each of the quasi 2D superconductors, which is calculated from specific heat data, and the influence of many body effects.  The calculated Pauli paramagnetic limits are then compared to critical field data for the title compound and other organic conductors.      Many of the examined quasi 2D superconductors, including the above organic superconductors and CeCoIn$_5$, exhibit upper critical fields that exceed their calculated $H_p$ suggesting unconventional superconductivity. We show that the high field low temperature state in $\kappa$-(BEDT-TTF)$_2$Cu(NCS)$_2$ is consistent with the Fulde Ferrell Larkin Ovchinnikov state. 
\end{abstract}


\maketitle

\section{Introduction}

Recently there has been interest in organic, heavy fermion, and pnictide superconductors exhibiting superconducting properties consistent with the exotic superconducting behavior\cite{singleton_symington00_jop_condmat, tanatar_suzuki03_synthmet, radovan_fortune03_nature, bianchi_movshovich03_prl, uji_shinagawa01_nature, BergkDemuer11PRB, KoutroulakisStewart10PRL, ChoKim11PRB} predicted by Fulde Ferell \cite{fulde_ferrell64_pr} and Larkin and Ovchinnikov,\cite{LarkinOvchinnikov65JETP} referred to here as the FFLO state.   This superconducting phase allows Cooper pairs to form with non zero momentum, and as a result has a spatially inhomogeneous order parameter. In this manuscript we describe measurements that show strong evidence for the existence of a FFLO state in the material $\kappa$-(BEDT-TTF)$_2$Cu(NCS)$_2$, referred to here as CuNCS.  The FFLO state lies above a phase line that is determined to be at the Pauli Limit by a semi-empirical calculation. 

One method to study superconductivity is to examine what happens when the superconducting material transitions from the superconducting state to the normal state.  Superconducting materials can be driven normal in one of two ways.  Either the temperature of the sample can be raised above the critical temperature ($T_c$), or an external magnetic field can be applied to the sample to raise it above the critical field ($H_{c2}$).  As an example, in a type II superconductor application of an external magnetic field causes vortices containing normal quasiparticles to form within the sample.  One explanation for how superconductivity is destroyed in this case is that as the magnetic field is increased, the density of the vortices increases until eventually they merge and drive the entire sample normal. 

At arbitrary orientations to the applied magnetic field in layered superconducting materials, vortices intersect the conducting planes and may form coupled pancake vortices.\cite{mansky_chaikin93_prl}  However, when layered superconductors are oriented with the conducting planes exactly parallel to the applied magnetic field, the magnetic field lines favor the less conducting material between the conducting layers and if the anisotropy of the conductivity is great enough the formation of traditional vortices is greatly suppressed, and the cordless Josephson vortices that do form are weakly pinned. It has been shown previously that the pinning force constant for parallel vortices moving between the layers in CuNCS is 500 times less than perpendicular vortices.\cite{mansky_chaikin94_prb} It is also explained in this previous reference that the Josephson vortices do not decrease the order parameter in the superconducting layers. 

Because vortices in the parallel orientation are no longer driving the sample normal, another mechanism must come into play to destroy the superconducting state.  The upper limit of superconductivity now occurs when the Zeeman spin splitting energy from the applied magnetic field approaches the value of the superconducting energy gap.  When this happens the Cooper pairs are broken and superconductivity is destroyed.  The field at which this happens is known as the Pauli paramagnetic limit. \cite{clogston62_prl} For an FFLO state to form in a superconductor, Pauli limiting must be more dominant than orbital limiting. Maki defined the parameter $\alpha_M = \sqrt(2)H_{c2}^o/H_p$, where $H_{c2}^o$ is the critical field at zero temperature, to measure the relative strengths of the two limiting mechanisms.\cite{maki66_pr}  He claimed that a material with $\alpha_M  > 1.8$ was a good candidate for finding the FFLO state.

For BCS superconductors the Pauli paramagnetic limit is given as
\begin{equation} 
H_p=\frac{\Delta}{2\sqrt{2}\mu_B}=1.83T_c,
\label{eq:hp1}
\end{equation} 
where $\Delta=1.76k_BT_c$ is the BCS energy gap.  However, many organic superconductors, such as CuNCS and 
$\beta''$-(BEDT-TTF)$_2$SF$_5$CH$_2$CF$_2$SO$_3$ (ET-SF5), and heavy fermion conductors, such as CeCoIn$_5$, do not  necessarily follow BCS theory.  Therefore, a more detailed method to find the Pauli paramagnetic limit must be used to take into account the deviation from BCS theory. In Section II we will describe our semi empirical Pauli Limit calculation. 

In Section III we will explain our experimental methods, and in particular give some details about our tunnel diode oscillator (TDO) penetration depth measurement system. 

We have studied CuNCS extensively over the past 15 years. It was initially our recent pulsed field data that allowed us to identify phase transitions in the H-T phase diagram that are consistent with an FFLO phase intervening between the traditional vortex state and the normal state at low temperatures as one moves up the field axis.  Given the clearer transitions in the recent data, we reanalyzed older data and found confirmation of our results in pulsed and dc magnetic field experiments that were done down to 50 mK.  The data for CuNCS comes from studies  of a number of different samples from different sample growers, and in different venues in pulsed and dc magnetic fields as described in Section IV. In all cases it shows consistent results. Given our confidence in these results, we compare the Pauli Limit found in the measurements with our semi empirical calculation of the Pauli Limit based on the gap energy as measured from specific heat.  The measured Pauli Limit agrees with our calculation, and with this confidence we apply the calculation to other organic superconductors in Section V. 

\section{Calculating $H_p$}

One aspect of organic and heavy fermion superconductors that deviates from BCS theory is the size of the energy gap in relation to $T_c$.  The standard relationships are that $\Delta=1.76k_BT_c$ for  BCS s-wave superconductors  and $\Delta=2.14k_BT_c$ for d-wave superconductors.  The theories are ideal and may not be correct for all classes of superconductors, so we have chosen to find the actual size of the energy gap by fitting the "$\alpha$-model" \cite{padamsee_neighbor73_joltp} to specific heat data.  Here $\alpha$ is the ratio of $k_BT_c$ and $\Delta$, 1.76 for standard BCS superconductors as mentioned above. The electronic specific heat below $T_c$ is used for this fit, not just the height of the specific heat jump. In general, if the specific heat data extends to lower temperatures, the quality of the fits is much better. With this method we were able to empirically determine $\alpha$ and thus the energy gap for all of our materials.\cite{tony}  (We note that this $\alpha$ is not related to Maki's $\alpha_M$ which we distinguish by the subscript M.) The resulting energy gap, using either an s-wave or d-wave fit, can then be used to calculate $H_p$ in a method based on a calculation by McKenzie.\cite{mckenzie99_arxiv} 

In addition to using an empirically determined energy gap, it is also important to include many-body effects.  These effects can be significant, as exhibited by the effective mass of the electron, which can be two to five times larger than the prediction from band-structure calculations.   Zuo et al.\cite{zuo_brooks00_prb} suggests reducing the energy gap by a factor of $g/g*$ which comes from the consideration of the Pauli Limit from within a Fermi liquid framework.  The ratio $g/g*$ can be experimentally determined from thermodynamic measurements or the spin-splitting of magnetic quantum oscillations. We can determine the effective mass and  $g/g*$  from Shubnikov -de Haas oscillations measured in the same experimental runs as our critical field measurements. Combining Eq. \ref{eq:hp1} to find the Pauli limit with the $\alpha$-model fit and including the many-body effects results in the semi empirical Pauli limit formula:
\begin{equation}
H_{P}=\left(\frac{g}{g^{*}}\right)\frac{\alpha k_{B}T_{c}}{\sqrt{2}\mu_{B}}.\label{eq:hp}\end{equation}
The fitted $\alpha$ values, $g/g*$ values and calculated $H_p$ for several organics superconductors and the heavy fermion superconductor CeCoIn$_5$ can be see in Table \ref{tab:calc_values}.

There has been much debate about the pairing symmetry of organic and heavy fermion superconductors. Rather than enter this debate, for each of our fits we tried both s and d wave versions of the theory to do the fits, and picked the best result. There was overwhelming evidence that the d-wave fits matched the data better for CuNCS. For the other materials the specific heat data was not as high quality and the relative quality of the  s and d wave fits was ambiguous.  In those cases we picked the d wave fits. Given some of the results in the literature \cite{analytis_ardavan06_prl} hybrid s + d wave models may need to be developed.

\begin{table*}
\caption{Experimentally determined $\alpha$ values, $g/g*$ values and calculated $H_p$. For these materials, and  NH4 in particular, that are truly Pauli limited over the entire temperature range, the slope ratio method for $H_p$ can give misleading results because the parallel critical field will ideally start with an infinite slope.\label{tab:calc_values}}
\begin{tabular}{|l | c | c | c|c | c | }
\hline
\hline
Material & $\alpha$ & $g/g*$ \cite{mckenzie99_arxiv} & $T_c$ & $H_p$& $\alpha_M$   \\
\hline
$\kappa$-(BEDT-TTF)$_2$Cu(NCS)$_2$ & 3.1 \cite{taylor_carrington07_prl} & 0.71 & 8.92 K & 20.67 T & 5.5   \\
\hline
$\beta''$-(BEDT-TTF)$_2$SF$_5$CH$_2$CF$_2$SO$_3$ & 2.0 \cite{wanka_hagel98_prb} & 1.0 & 4.7 K & 9.9 T & 3.9  \\
\hline
$\alpha$-(ET)$_2$NH$_4$Hg(SCN)$_4$ & 1.76 \cite{elsinger_wosnitza00_prl} & 1.16 & 0.93 K & 2.0 T & 9.9 \\
\hline
$\lambda$-(BETS)$_2$GaCl$_4$ & 2.4 \cite{ishizaki_uozaki03_synthmet} & 1.0 & 4.9 K & 12.38 T & 13.2  \\
\hline
CeCoIn$_5$ & 2.32 \cite{park_greene05_prb} & 1.56 & 2.3 K & 8.78 T & 6.1 \\
\hline
\hline
\end{tabular}
\end{table*}

\section{Experimental Methods}
To measure the possible phase changes in our superconducting systems, we choose a penetration depth measurement.  We needed a probe that would yield information even in the superconducting state, which, for example, rules out transport. The field versus temperature data presented in this manuscript was taken using a tunnel diode oscillator (TDO) technique.  The TDO technique is a  contactless measurement, which allows for the study of very small samples and reduces stress on the samples.  The TDO circuit consists of a coil attached to a resonant circuit \cite{coffey_bayindir00_rsi} that is driven by a tunnel diode. The sample is placed in the coil and changes of the rf penetration into the sample change the frequency of the resonant circuit. We have used oscillators from 25 - 1050 MHz with frequency resolution of 5 - 200 Hz depending on the venue and whether the experiment was conducted in pulsed or dc magnetic fields.  

In most of the experiments the samples were held at a constant temperature while the magnetic field was swept either by pulsing a magnet or by changing the dc field.  As the external magnetic field is changed, the frequency of the resonant circuit changes.  At times we swept the temperature at zero field to find $T_c$. In these zero field experiments vortices do not play a role. The change in frequency of the TDO is proportional to the change in rf penetration depth of the sample. \cite{coffey_bayindir00_rsi}  The change in the rf penetration depth is a complicated mix of skin depth, London penetration depth, and coupling to vortices, where the mixture of these mechanisms depends on the orientation of the sample to the applied magnetic field, and the state of the sample as it changes from a conductor to a superconductor. Although it is difficult to separate all the causes, the penetration depth is a very sensitive probe of the sample's electronic structure. Detailed analyses of the interaction of rf fields with superconducting samples can be found in a number of papers, \cite{CoffeyClem1991PRL, ProzorovGiannetta2000PRB} and particular examples of other experiments where the rf penetration depth is measured with a TDO can be found in the literature.\cite{CoffeyMartin10PRB, mielke_singleton01_jop, FletcherCarrington2007PRL,  HashimotoKasahara2012PRL}  

The pulsed field data was taken at Clark University's Pulsed Field Magnet Laboratory.  In this laboratory low temperatures are achieved using a combination $^3$He/$^4$He cryostat, which allows the sample to be cooled down to  400 mK.  The applied magnetic field is provided by our 42 and 50 T pulsed field magnets that reach maximum field at 25 and 11 ms respectively.  The dc data was taken at the National High Magnetic Field Laboratory (NHMFL) in Tallahassee, FL.  At the NHMFL the sample was temperature controlled using a portable dilution refrigerator in the range of 47 - 600 mK and a 33 T dc magnet was used to apply an external magnetic field.

In both venues the sample and coil were mounted on a rotating platform that was controlled by a micrometer at the top of the cryostat. The angular resolution in all the experiments was better than 0.1 degree. 

\section{Experimental Results $\kappa$-(BEDT-TTF)$_2$Cu(NCS)$_2$}
\begin{figure}
\includegraphics[scale=0.45]{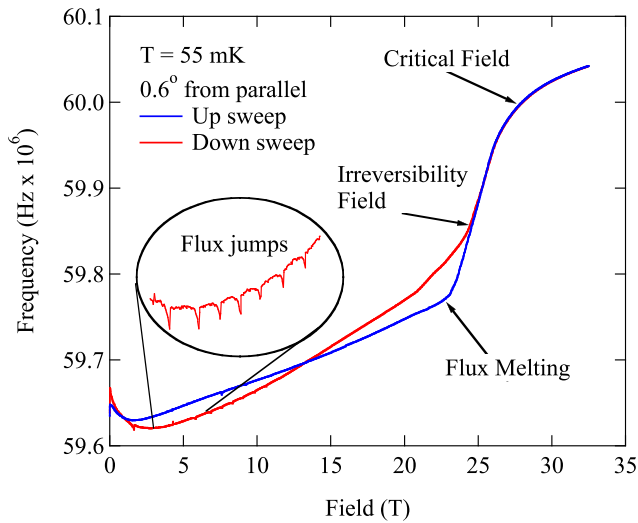}
\caption{Frequency versus field data for CuNCS taken slightly off parallel.  The upsweep is in blue and the downsweep is in red.  The melting transition ($H_m$), the irreversibly point ($H_{irr}$) and the upper critical field ($H_{C2}$) are indicated with arrows.\label{cap:data_vortex}}
\end{figure}

In order to claim that the cause of the superconducting to normal state transition is dominated by reaching the Pauli Limit and not vortex effects, it is necessary to show that orienting layered superconducting samples with their conducting planes parallel to the applied magnetic field suppresses traditional vortex formation.  To illustrate the effects of vortices we have taken high resolution TDO data as a function of dc magnetic field for temperatures from 55 to 600 mK and at orientations of the conducting planes spanning from parallel to perpendicular with respect to the magnetic field. In these experiments  the temperature was held constant and the magnetic field was swept from 0 to 33 T and back again for different sample orientations. The data show a number of features which we have identified in Fig.\ref{cap:data_vortex}, such as the critical field, $H_c2$,  the vortex melting transition ($H_m$), the irreversibility transition ($H_{irr}$), and  at the lowest temperatures,  flux jumps.  

We have collected data roughly three degrees on either side of parallel.  Looking at Fig. \ref{cap:data_vortex} the irreversibility line is obvious given that it is defined by the point where the hysteresis between the up and down sweeps ends.     $H_m$, where magnetic field becomes strong enough that the quasi-2D vortex lattice is able to de-pin from the conducting layer, can be seen as a kink in the data trace if the sample orientation is greater than $0.3^\circ$ away from parallel.  When the sample planes are almost exactly parallel (black trace) to the magnetic field, the melting transition and the irreversibility line are no longer seen.

\begin{figure}
\includegraphics[scale=0.450]{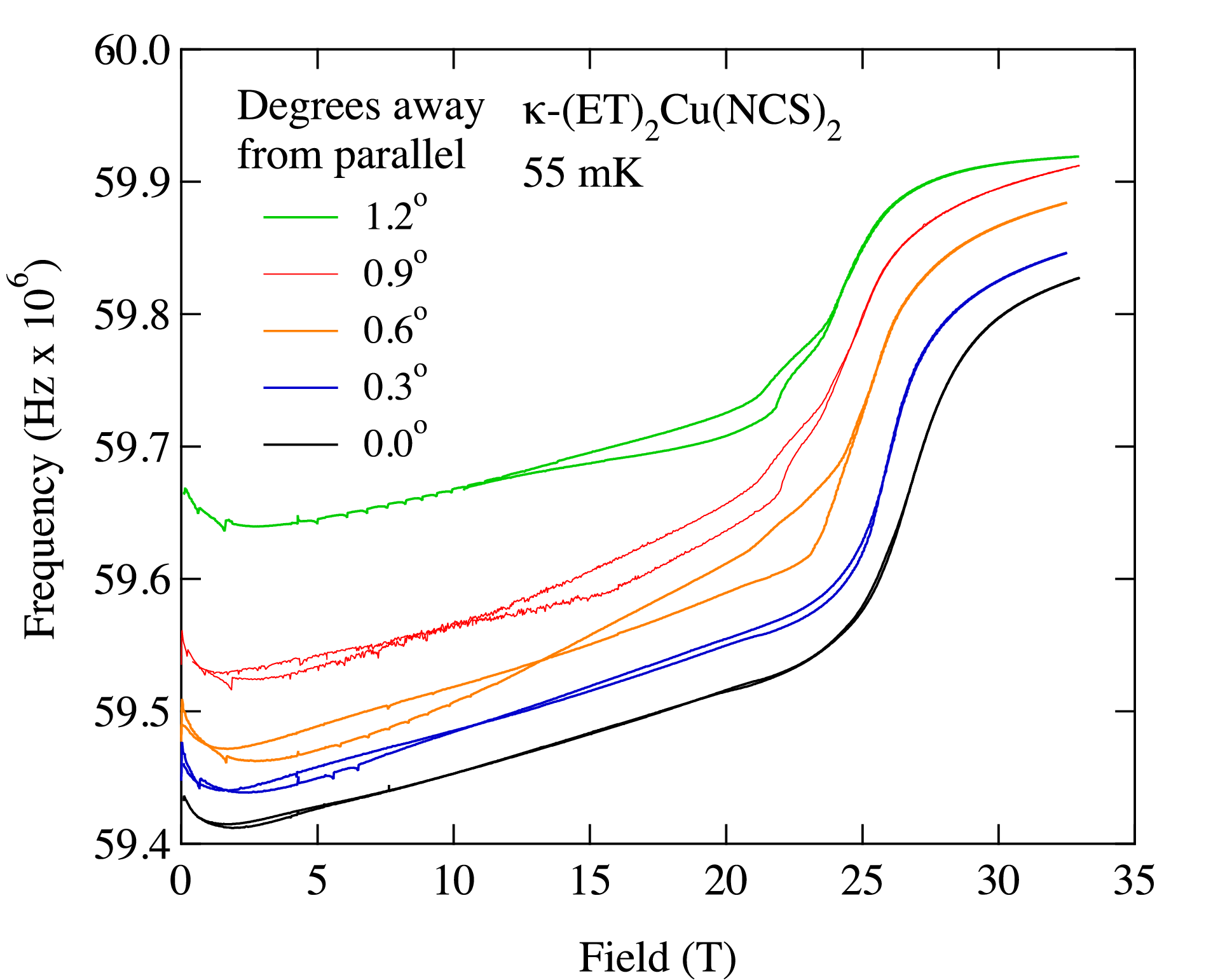}
\caption{The angle dependence of the penetration depth near the parallel orientation for CuNCS. The lowest trace is exactly parallel, or $90^\circ$ in our absolute coordinates.  The other traces are in order of increasing angle.  The traces are vertically shifted to aid visualization. It is remarkable how all the vortex details are absent at the exactly parallel orientation.  \label{nearParallel}}
\end{figure}

It is a remarkable and important observation that the features related to the vortices, the melting transition, the irreversibly line, and the flux jumps, all are absent when the sample is aligned within 0.3 degrees of the parallel orientation as shown in Fig. \ref{nearParallel}.   The absence of these features is due to the ability of the vortices to form between the most conducting layers where their influence on the bulk superconducting state is minimized.\cite{mansky_chaikin94_prb} The up and down sweeps in the parallel orientation are remarkable because they show no flux jumps, no hysteresis, and no obvious kinks. In the limit of poor conductivity between the layers, the vortices are replaced by cordless Josephson vortices that can slip in from the edges of the sample, and vortices are no longer important to understand the physics of the superconducting state.  In this orientation the superconductivity is very weakly limited by the orbital properties of the vortices, and the superconductivity becomes Pauli Limited.

\begin{figure}
\includegraphics[scale=0.45]{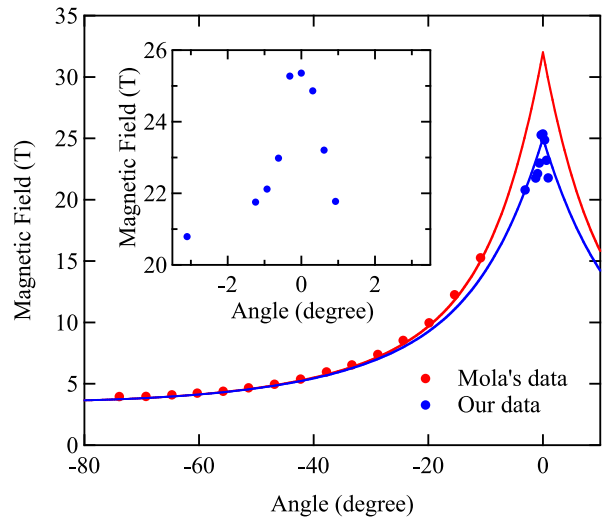}
\caption{The melting line is indicated as a function of angle at 0.55 K.  The data is compared to Mola's data taken up to an angle of $75^\circ$ and the Tinkham formula, also found in Mola's paper. Our data suggests a much lower field for the extrapolation of the melting transition at the parallel orientation. \label{cap:melt}}
\end{figure}

 The melting of the vortex lattice is identified by comparing to previous data by Mola et al,\cite{mola_hill01_prl} and noticing that the transition goes to higher fields as the sample approaches parallel.  The peak of this transition near parallel is at a higher field than predicted by Mola if we use the Tinkham formula as they did to extrapolate their fit as shown in Fig.~\ref{cap:melt}.  The Tinkham formula only has two parameters, $H_{c2}$ parallel and perpendicular, so for our fit we just used the extremes of our data, as seen in Fig.~\ref{cap:melt}. The differences between the two fitted melting lines could be due to differences in the density of defects in the samples, which may depend on details of the synthesis process. We note that within $0.5^\circ$ of parallel we claim in Section V that we have entered the FFLO state and the Tinkham formula should not apply for the melting line or irreversibility line.  

\begin{figure}
\includegraphics[scale=0.45]{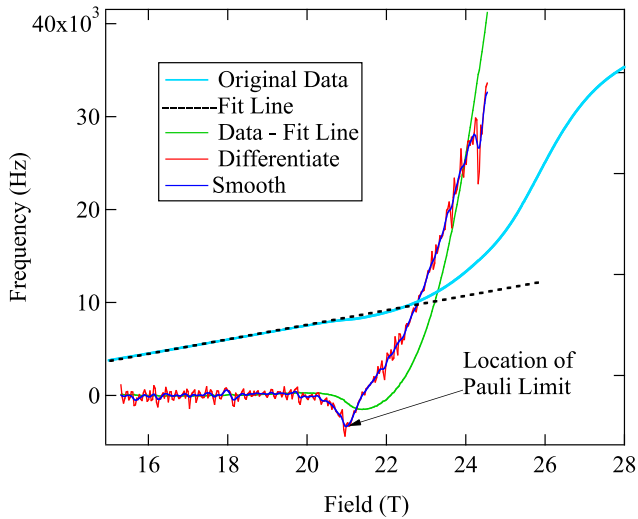}
\caption{Parallel data for CuNCS  at 630 mK is shown in light blue and the black dashed line is a fit to the initial part of the sweep and used as a background.  This background is subtracted off leaving the green trace, with its derivative shown in red.  The smoothed derivative is shown in dark blue. \label{cap:orig}}
\end{figure}

\begin{figure}
\includegraphics[scale=0.45]{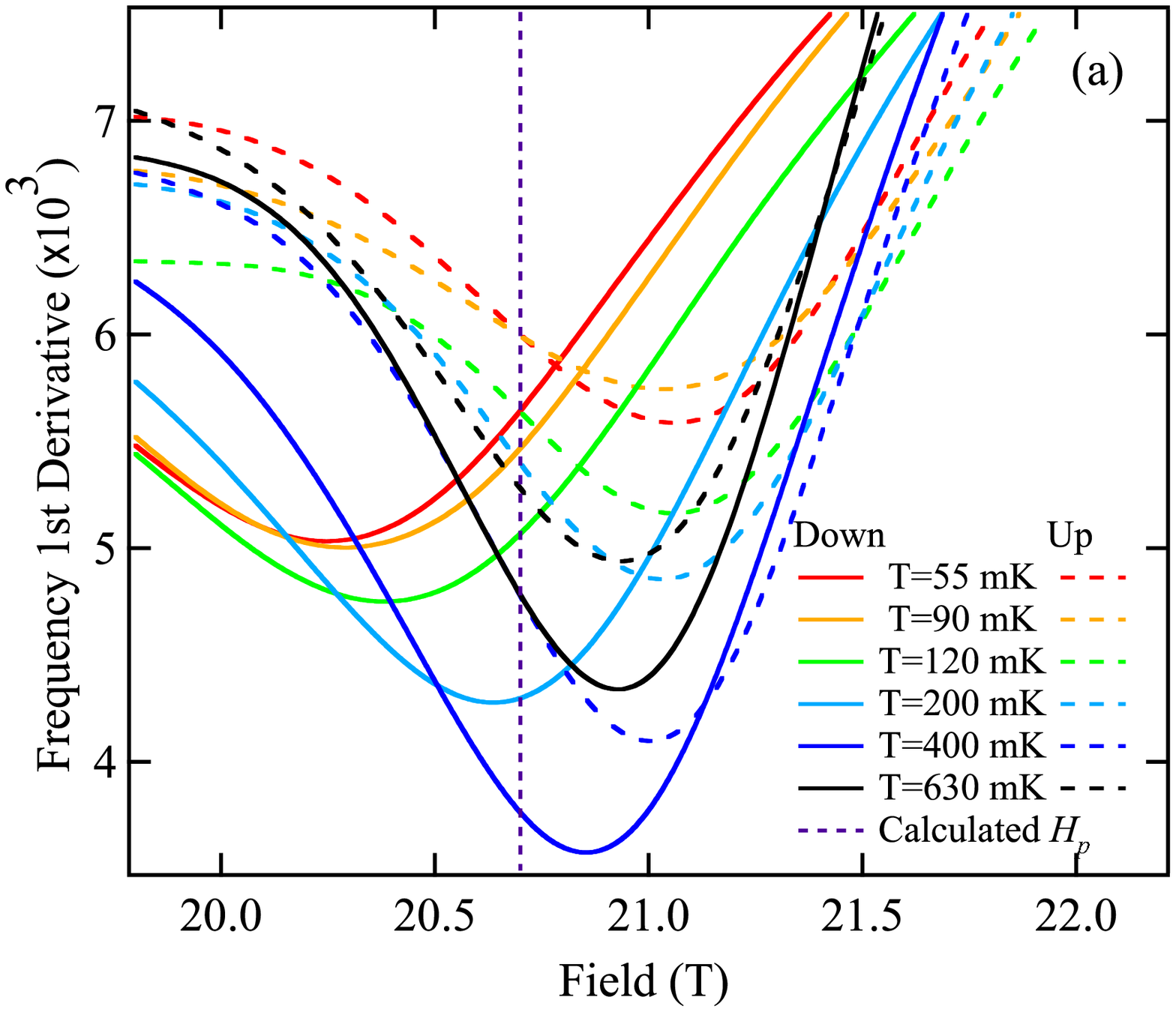}
\includegraphics[scale=0.45]{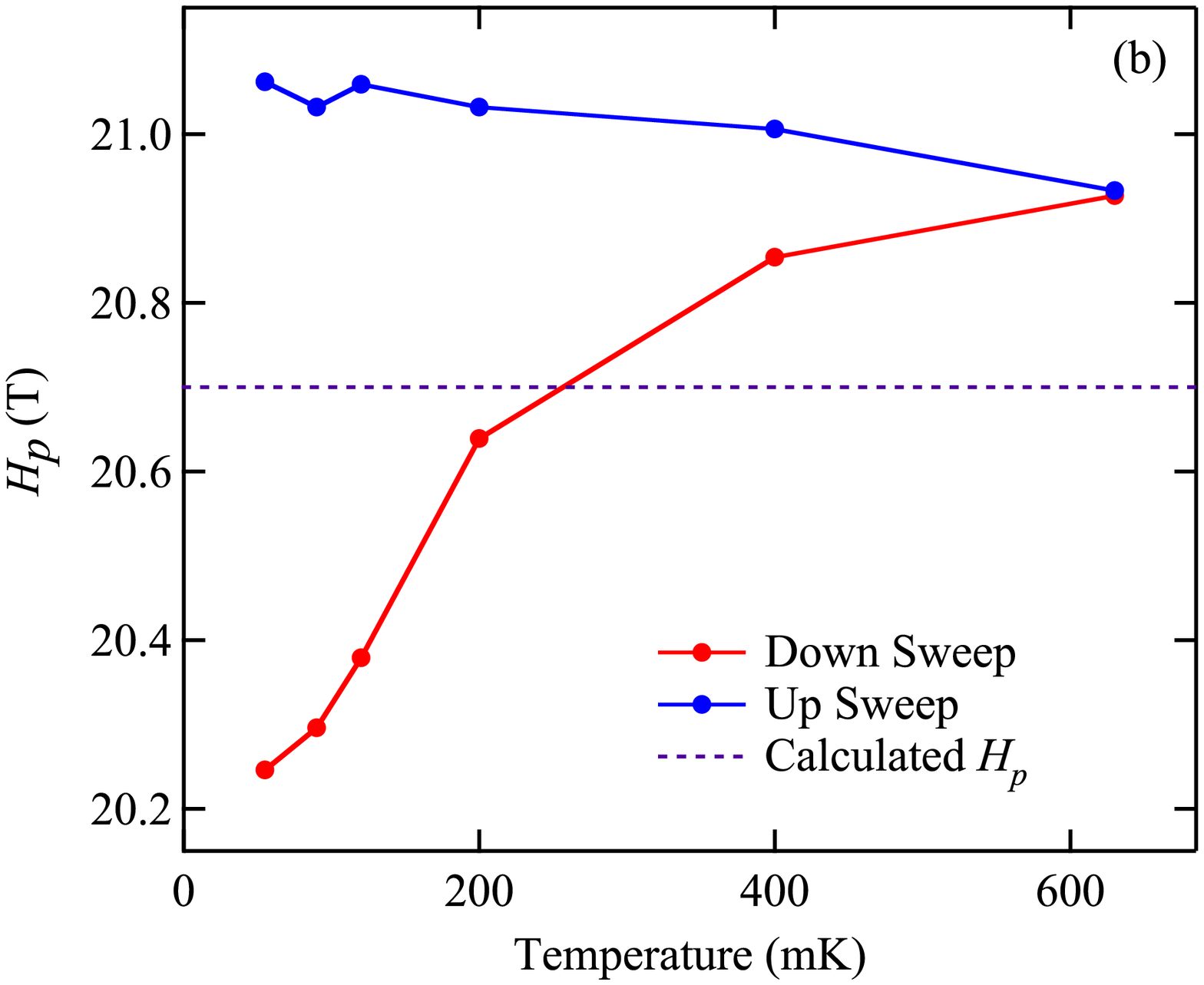}
\caption{(a) The first derivative up and downsweeps at the parallel orientation for CuNCS for several different temperatures. (b) The feature seen in the data in Fig.\ref{cap:orig} is tracked and should be compared to the calculated $H_p$ of 20.7 T in both the upsweep and downsweep data for a range of temperatures. The widening gap between the up and down sweeps shows some hysteresis in the measurement and suggests a possible first order transition. \label{cap:hp}}
\end{figure}

Now that the vortex effects are accounted for, there is a remaining feature in the data.  In order to be able to visualize this feature better, a straight line was fit to the initial part of the parallel trace, up to about 20 T, (Fig. \ref{cap:orig}), and then subtracted from the original data ("Data - Fit Line" in green). This subtraction eliminates the background frequency change due to the increasing magnetic field.  We then take the derivative of this subtracted data (Fig. \ref{cap:orig}  in red).  This derivative is then smoothed (Fig. \ref{cap:orig} blue).  What remains is a feature that is located close in field to our calculated $H_p$. This possible signature of $H_p$ is seen at several different temperatures (Fig. \ref{cap:hp}) and is seen in both the upsweep and downsweep of the magnetic field.  The dashed black line indicates the calculated $H_p$.

\begin{figure}
\includegraphics[scale=0.45]{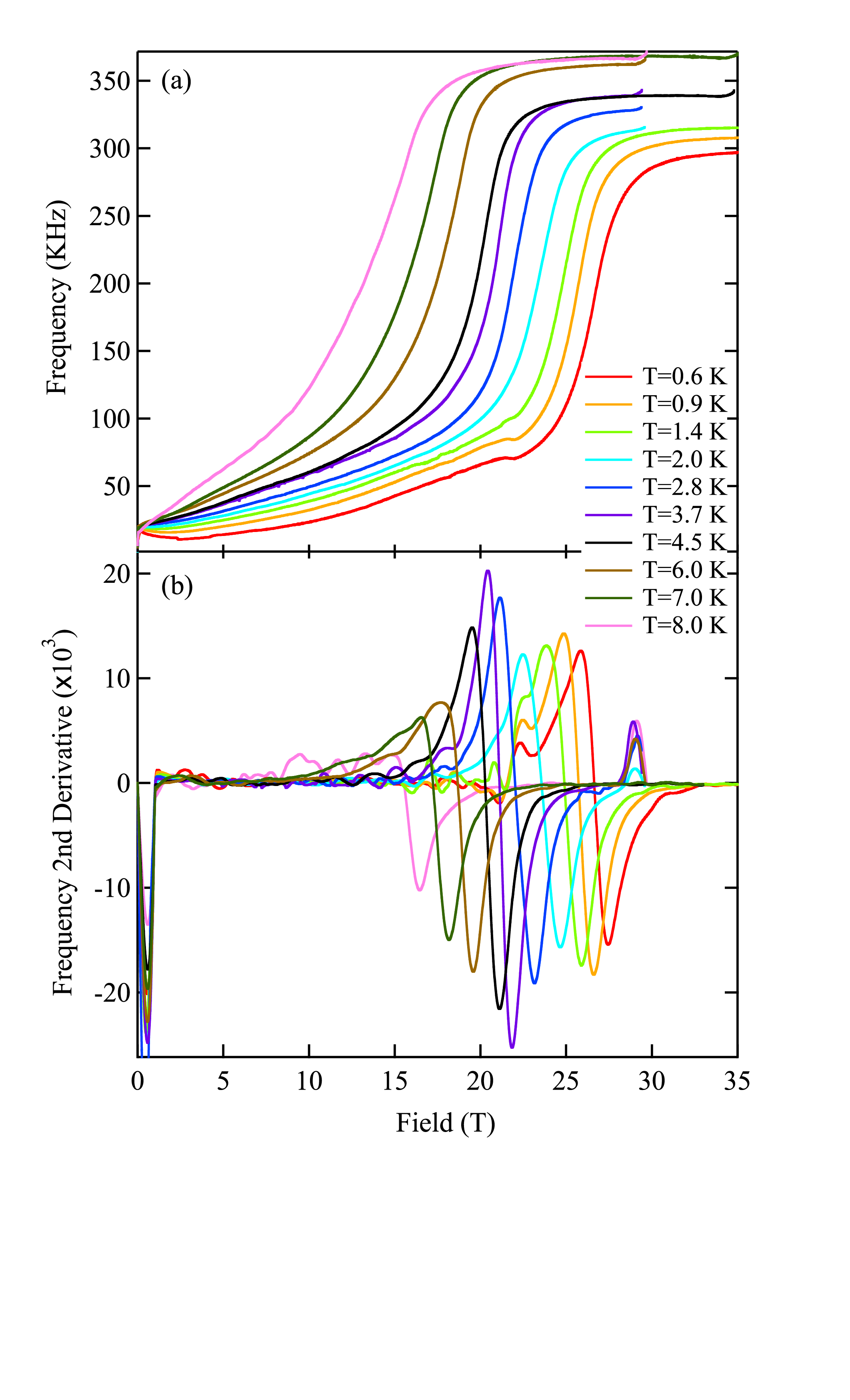}
\caption{(a) Field sweeps as a function of temperature in CuNCS-d8.   (b) From the second derivative the transitions can be picked out.  We choose the maximum negative curvature for the location of $H_{C2}$, and the local lower maximum positive curvature (small bump) of the lower transition to locate the Pauli Limit.  It is clear from these second derivative traces that $H_{C2}$ is function of temperature, but $H_m$ is not nearly as much. 
\label{cap:NCSTemper}}
\end{figure}

The feature is not present if the orientation of the sample is moved off of parallel by more than $0.3^\circ$, suggesting that the presence of vortices, and vortex limiting, causes the phase line to disappear.  We note that in a nonparallel magnetic field $H_p$ still exists, but not the FFLO state. As a function of temperature, we do not expect the field value of the transition to move very much, and in this venue we could only cover temperatures up to 600 mK.  In that small range the field of the transition changes by only about 1 T as seen in Fig.  \ref{cap:hp}.  These facts are consistent with a phase line separating a vortex superconducting state with an FFLO state. The FFLO state should have nodes in the order parameter, suggesting in the first approximation that the fewer Cooper pairs will result in greater rf penetration.\cite{martin_agosta05_prb} We also note that the phase lines showed some hysteresis between the up and down sweep suggesting a first order transition. Finding this phase line was difficult given all of the other features in the data, but as we show in Fig. \ref{cap:NCSTemper}, we found a similar feature in more recent pulsed magnetic field data using samples from a different sample grower, which gives us great confidence in the location of this phase line. Although previous reports have claimed to find this same phase line\cite{singleton_symington00_jop_condmat, lortz_wang07_prl} the slope of the earlier data is not near zero as one would expect for the Pauli limit, or consistent with the theory shown below.  Only the more recent data of Bergk\cite{BergkDemuer11PRB}  show data consistent with ours and the theory, at least above 2 K, as does recent NMR data.\cite{WrightGreen2011PRL}

\begin{figure}
\includegraphics[scale=0.45]{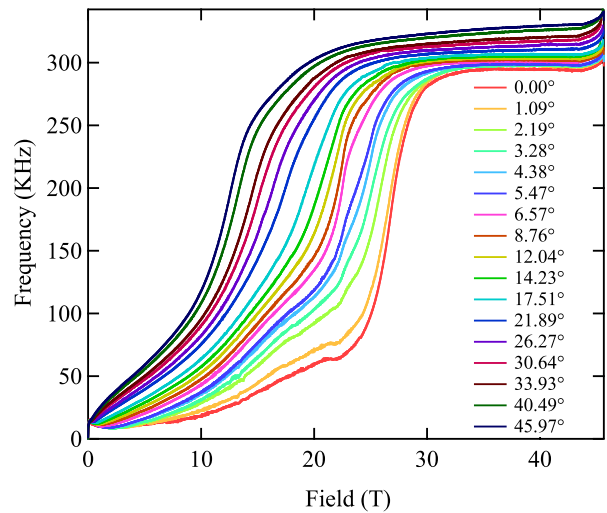}
\caption{The angle dependence of the penetration depth near the parallel orientation for CuNCS-d8 in pulsed fields. Compare to Fig.\ref{nearParallel}. The lowest trace is exactly parallel, or $0.0^\circ$.  
\label{cap:angleSweepPF}}
\end{figure}

For the more recent measurements in pulsed fields, shown in Figs.~\ref{cap:NCSTemper} and \ref{cap:angleSweepPF},  a slightly different sample was used, CuNCS-d8, where some of the hydrogens have been replaced with deuterium.\cite{BiggsKlehe02ConMat}  A clear bump at 22 T, a similar field to the previous FFLO transition, is seen.  Second derivatives make it easy to locate the position of all of these transitions and other features.  Similar to the first set of data, the bump goes away if the angle is not within  $2.0^\circ$ of parallel. In this pulsed field apparatus we were are able to take data at temperatures from 0.5 to 8.5 K and get a more complete phase diagram.    Looking back over older sets of data we have been able to locate the same feature near the Pauli limit, buried in noise. 

Using combinations of older\cite{Bayindir2001723, agosta_coffey02_ijompb} and newer data allows us to create a very complete phase diagram and increase our confidence in the accuracy of the data.  To make sure we are consistent we have reanalyzed the recent and past data using the same method of calculating second derivatives and using the location of the maximum positive or negative curvature of the rf penetration data to find the FFLO tradition and $H_{C2}$ respectively. This same analysis also locates possible other higher order FFLO transitions. We do expect a feature in the data when the vortex state is replaced by the FFLO state. The simplest reason is that the FFLO state has nodes in the order parameter, and on average fewer cooper pairs.\cite{martin_agosta05_prb}  Therefore, the London Penetration depth should change. There is also a possibility that the TDO is sensitive to the free quasi particles that exist at the nodes of the order parameter, although because we do not know the geometry of this node structure, it is difficult to predict the contribution of the quasiparticles to the penetration depth. 

\begin{figure}
\includegraphics[scale=0.45]{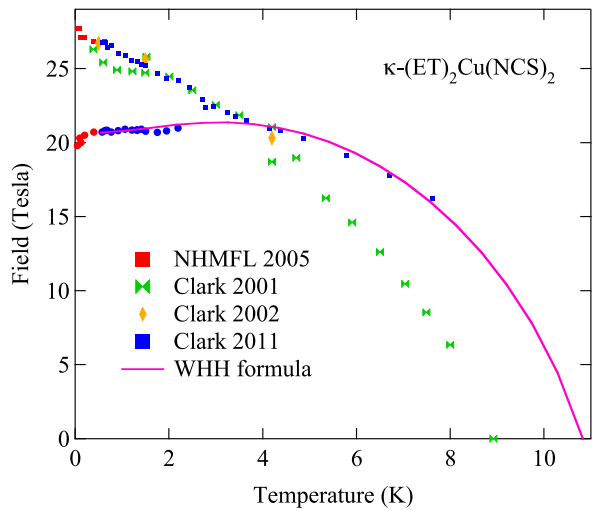}
\caption{The phase diagram for CuNCS oriented exactly parallel to the magnetic field.  The data comes from a number of samples and experiments in pulsed and dc fields. One of the samples is deuterated, which increases the energy gap,\cite{BiggsKlehe02ConMat} and should augment both the Pauli limit and the $H_{C2}$. The data is scaled for that difference.  The dark blue line is the WHH fit discussed in more detail in the main text.   \label{cap:phase}}
\end{figure}

The resulting phase diagram is shown in Fig. \ref{cap:phase}.  There is a clear $H_{C2}$  and a strong transition near our calculated Pauli Limit.  The most recent pulsed field data would  show slightly higher transitions, consistent with the larger energy gap of the deuterated compound, but we have scaled that data downwards accordingly to place it on the same plot. The consistency of the transitions across samples, sample growers, and experimental venues gives us confidence that the phase diagram is correct.  The shape of the phase diagram, and the location of the first phase transition near the calculated Pauli Limit suggests strongly that we have correctly calculated the Pauli limit, and can identify the upper phase  as a a different phase.  The position and shape of the phase on the H-T diagram is consistent with an FFLO state.  The Clark 2001 and NHMFL 2005 data show upturns in the data at the lowest temperatures, consistent with theoretical predictions for FFLO.\cite{HouzetMeurdesoif99PhysC} The lower set of high temperature data labeled Clark 2001 is not consistent with our expectations, but at high temperatures the steep phase line makes field sweeps a very insensitive measurement of $H_{C2}$, and the old data was much less clear than the 2011 data. 

It would be ideal to fit the data to recent theories that calculate the shape of the phase lines that form the FFLO state, but the theories are numerous and we hope to have that problem sorted out for our next paper.  Nevertheless, it is possible the use the WHH  formula,\cite{werthamer_helfand66_pr} a well established theory, to fit the border of the vortex state. Although the WHH formula is usually used to fit $H_{C2}$ in traditional type II superconductors, it should work just as well to fit the limit of the traditional vortex state in a superconductor with a FFLO state.  We need to introduce one more parameter to use WHH, the spin orbit scattering parameter $\lambda_{so}$. We will use a unitless version of $\lambda_{so}$ following WHH, and defined as 
\begin{equation}
\lambda_{so} = \frac{h}{3k_BT_c\pi\tau_{so}},\label{eq:so}
\end{equation}
where $\tau_{so}$ is the spin orbit scattering time and $h$ is Plank's constant.  Spin-orbit scattering is important because at high fields it lowers the probability of depairing Cooper pairs, thus raising the Pauli Limit and consequently $H_{c2}$. For similar reasons spin-orbit scattering makes the FFLO state less likely.  
The result of the WHH fit is shown in Fig.  \ref{cap:phase} along with all of the data.  The fit results $H_p = 19.2, \alpha_M = 4.1$ and $\lambda_{so}$ = 0.077 are all consistent with other estimates of the same parameters in this paper.  This application of the WHH formula is a solid confirmation of the somewhat subtle phase line between the vortex and FFLO superconducting phases. 

\section{Discussion}

Given the success with the material CuNCS, we now turn to other quasi 2D superconductors we have studied. ET-SF5\cite{ChoSmith09PRB} and ($\lambda$-BETS)\cite{ConiglioWinter11PRB} have phase diagrams similar to CuNCS, and $\alpha$-(ET)$_2$NH$_4$Hg(SCN)$_4$ \cite{CoffeyMartin10PRB}  (NH4) is an example of a clean but fully Pauli limited superconductor. From the references for the above materials we find the experimental  $H_p$ values 10.5, 9.5, and 2.15 for ET-SF5,\cite{ChoSmith09PRB} $\lambda$-BETS, \cite{ConiglioWinter11PRB} and NH4 \cite{CoffeyMartin10PRB} respectively. The measured and calculated $H_p$s are summarized in Table \ref{tab:Hp}. The first three measured $H_p$ values are all within 5 \% of the calculated values.  The agreement of the other two values is not as good, but the $\lambda$-BETS specific heat data is not of the highest quality, and CeCoIn$_5$ has a phase diagram that does not follow the expected Pauli limiting shape (see KLB \cite{klemm_luther75_prb} for the theory and look at the experimental data here\cite{KoutroulakisStewart10PRL}) unlike the organic samples.  


\begin{table*}
\caption{The calculated $H_p$ from the method of Sec. II and the measured $H_p$ from the critical field measurements are shown in the first two columns of this table.  The next column is the Maki parameter as found from the fits of the critical field (or extent of the vortex state) using the WHH formula. This data comes from the papers cited in the main text for each of the materials.  The energy gap is calculated from $\alpha$ in Table I and $T_c$. Finally $t^*$, the superconducting coherence length $\xi$, the mean free path mfp, and r =  mfp/$\xi$ are listed.  \label{tab:Hp} }
\begin{tabular}{| l | c |c | c | c  | c  | c | c | c | }
\hline
\hline
Material & Calculated $H_p$ & Meas. $H_p$  & $\alpha_M$ WHH & $\Delta$ & $t^*$ & $\xi$ & mfp & r \\
\hline
$\kappa$-(BEDT-TTF)$_2$Cu(NCS)$_2$ & 20.67 T & 20.7 T  & 8.4  & $4.28x10^{-22}$ J & 0.38 & 74 & 1513 & 20 \\
\hline
$\beta''$-(BEDT-TTF)$_2$SF$_5$CH$_2$CF$_2$SO$_3$  & 9.9 T & 10.5 T & - &  $1.42x10^{-22}$ J & 0.24 & 158 & 520 & 3.3 \\
\hline
$\alpha$-(ET)$_2$NH$_4$Hg(SCN)$_4$ &  2.0 T & 2.15 T & 7.1 &  $2.26x10^{-23}$ J  & - & 628 & 600 & 0.96\\
\hline
$\lambda$-(BETS)$_2$GaCl$_4$ &  12.38 T & 9.5 T  & 12.9 &  $1.62x10^{-22}$ J & 0.35 & 108 & 170 & 4.0 \\
\hline
CeCoIn$_5$ &  8.78 T  & 10.0 T  & -  &  $7.36x10^{-23}$ J & 0.13 & 58 & 810 & 14\\
\hline
\hline
\end{tabular}
\end{table*}

The data from this class of highly anisotropic clean superconductors consistently show new phase lines that could be due to a FFLO state, that are present in their H-T phase diagrams.  In contrast, NH4 is a good example of a clean anisotropic material that because of its weak superconductivity has a very long coherence length, and cannot support the extended wave function necessary for the long range order of the FFLO state.  Another good counter example is CuNCS under pressure.\cite{martin_agosta05_joltp}  The critical temperature is reduced by the addition of pressure, and a cursory look at the phase diagram suggests that the FFLO state is suppressed, producing a phase diagram similar to NH4.  

One question we can ask, now that we have a number of potential FFLO phase diagrams, is if they are consistent with theory.  There are many theory papers in this field that have been published over the last 40+ years, and we can get reasonable fits to some of the theories\cite{Shimahara1994PRB, BuzdinBrison1996EPL, houzet_buzdin02_prl} as shown in Coniglio,\cite{Coniglio11} although due to the complexity of the fits we will leave that discussion to a separate paper.  One particular issue we can address immediately is the location of the tricritical point. Many theories show that the location of the tricritical point, $t^* = T^*/T_c$,  should equal  0.56. \cite{Shimahara1994PRB, HouzetBuzdin2001PRB}  As shown in Table \ref{tab:Hp}, the tricritical points that we measure range between 0.29 and 0.38.  These values are lower than the value predicted by theory, and that difference begs the question of what is depressing the value of $t^*$. There are two groups of parameters that we might associate with the value of $t^*$, dissipative parameters such as scattering, or intrinsic structure such as the Fermi surface.  Shimahara argues  that the Fermi surface should not determine the position of $t^*$, but he does say that details of the Fermi surface's ability to nest could determine how robust the FFLO state is. \cite{Shimahara1994PRB}  We will examine dissipative parameters such as the scattering time and indirectly the mean free path, in addition to spin-orbit scattering.   A short mean free path would limit the coherence of a long range wave function, and in this case it is the the q-vector of the FFLO state that sets the long range distance periodicity of the wave function. 

To get an estimate of the scale of the FFLO periodicity, the distance between nodes, we calculate the q vector wavelength, $l_q$,  using the relationship $l_q = h/q_0$. From Shimahara\cite{Shimahara1994PRB} we find the formula $q_0 = 2\Delta_0/v_f$, where $\Delta_0$ is the superconducting energy gap at zero temperature, and $v_f$ is the Fermi velocity.  The wave vector $q$ gets larger as the temperature rises, so the $ l_q$  that we calculate at zero temperature, which is  $~340\AA$ for CuNCS, is a minimum value, and well below the mean free path according to Table \ref{tab:Hp}. 
 
In Table \ref{tab:Hp} we list $t^*$ and parameters such as the scattering time, mean free path, and the coupling constant $\alpha$ for a number of materials.  We note that determining the mean free path in organic conductors can be subtle, because the tradition measure of the Dingle temperature from quantum oscillations may not be valid.\cite{singleton_mielke03_jop_condmat}  Therefore, the scattering time for CuNCS comes from AMRO measurements.\cite{SingletonGoddard2007PRL}   The results in Table \ref{tab:Hp} for CuNCS and BETS  show that $t^*$ does not correlate with mean free path, with the caveat that the BETS mean free path was measured on a separate sample using SdH oscillations.  However the separate BETS sample did come from a similar batch to the ones studied for FFLO, and therefor the same sample grower, and given that the quantum oscillations did not start until 35 T, it is reasonable to assume that it does have a shorter mean free path than the CuNCS samples where the oscillations start at 12 T.  

Now we consider if the more intrinsic properties such as the energy gap or shape of the Fermi surface may be important to determine $t^*$.   It is important to point out that ET-SF5 has a highly elliptical 2D Fermi surface cross section\cite{WosnitzaWanka1999SM} unlike the other two organic conductors that we have studied where the Fermi surface cylinders are more round.\cite{GoddardBlundell2004PRB}  Given the earlier reference to the discussion by Shimahara\cite{Shimahara1994PRB} about degrees of Fermi surface nesting, the effects of Fermi Surface shape and nesting are also worth more study, and may have some influence on the small  $t^*$ for ET-SF5.
 
From the data we have so far, the energy gap and the size of the coherence length correlate best with the position of the tricritical point.  We note that ET-SF5 has the lowest $t^*$ and the smallest superconducting energy gap of the organic superconductors.  This relationship is a possible place to look for a mechanism that determines $t^*$. As described in some theory papers, and we cite one as an example,\cite{HouzetBuzdin2001PRB} to the extent that orbital effects need to be considered, either due to the degree of anisotropy, or the size of the effective mass of the Cooper pairs, the tricritical point will be effected.  In the case of the organic conductors considered in this manuscript, the degree of anisotropy is similar, but the spatial scale of the vortices is set by the the coherence length, which is in turn set by a combination of the size of the energy gap and effective mass.  This combination may be the key to understanding $t^*$.

In summary we have found phase diagrams in anisotropic superconductors that are consistent in shape and material parameters with predictions for an inhomogenious superconducting state.  We have shown that there is a high probability that the FFLO state exists at low temperatures and high magnetic fields in these superconductors.  We have begun the discussion of what determines the details of the extent of the FFLO state.

\begin{acknowledgments}
We thank John Singleton and Anthony Carrington for useful discussions, and we acknowledge support from DOE grant \#ER46214. In addition, a portion of the high magnetic field work was performed at the National High Magnetic Field Laboratory, which is supported by National Science Foundation Cooperative Agreement No. DMR-0654118, the State of Florida, and the U.S. Department of Energy,  and work at Argonne, a U.S. Department of Energy Office of Science laboratory, is supported under Contract No. DE-AC02-06CH11357.
\end{acknowledgments}

\bibliographystyle{apsrev4-1}
\bibliography{082608_bibfileb}

\end{document}